\documentclass[seceq]{ptptex}

\usepackage{graphicx}

\title{
High field properties of geometrically frustrated magnets}

\author{
M. E. \textsc{Zhitomirsky}$^1$ and Hirokazu \textsc{Tsunetsugu}$^2$}%
\inst{
$^1$Commissariat \`a l'Energie Atomique, DSM/DRFMC/SPSMS, \\ 38054 Grenoble, 
France, \\
$^2$Yukawa Institute for Theoretical Physics, Kyoto University, \\
Kyoto 606-8502, Japan}

\recdate{June 14, 2005}

\abst{         
Above the saturation field, geometrically frustrated quantum antiferromagnets 
have dispersionless low-energy branches of excitations  
corresponding to localized spin-flip modes.
Transition into a partially magnetized state
occurs via condensation of an infinite number of degrees of freedom.
The ground state below the phase transition is a magnon crystal,
which breaks only translational symmetry and preserves spin-rotations
about the field direction.
We give a detailed review of recent works
on physics of such phase transitions and present 
further theoretical developments.
Specifically, the low-energy 
degrees of freedom of a spin-1/2 kagom\'e antiferromagnet 
are mapped to a hard hexagon gas 
on a triangular lattice. 
Such a mapping allows to obtain a quantitative description
of the magnetothermodynamics of a quantum kagom\'e antiferromagnet
from the exact solution for a hard hexagon gas.
In particular, we find the exact critical behavior at the 
transition into a magnon crystal state, the universal value of the entropy
at the saturation field, and the position of peaks in temperature- and field-dependence 
of the specific heat.
Analogous mapping is presented for the sawtooth chain, which
is mapped onto a model of classical hard dimers on a chain.
The finite macroscopic entropies of geometrically frustrated magnets 
at the saturation field lead to a large magnetocaloric effect.
}

\begin{document}

\maketitle

\section{Introduction}

Theoretical investigation of finite field 
properties of geometrically frustrated magnets 
is a new rapidly developing direction of research, which 
is to a large extent stimulated by the experimental studies.
There are now a few examples of frustrated magnetic compounds such as pyrochlore 
Gd$_2$Ti$_2$O$_7$, \cite{ramirez02} garnet 
Gd$_3$Ga$_5$O$_{12}$, \cite{hov} 
and spinel CdCr$_2$O$_4$, \cite{takagi} for 
which full or partial reconstruction
of the phase diagram in magnetic field has been done.
From theoretical point of view, an applied magnetic field 
competes with antiparallel alignment of spins favored by antiferromagnetic
exchange interaction and creates an extra source of frustration. 
This may lead to various interesting phenomena
at high magnetic fields such as exotic quantum ground states, 
novel types of the critical behavior, the magnetization jumps and
plateaus and so on.
Here we give a detailed overview and present further development
of the recent theoretical works on
a universal behavior of geometrically frustrated 
magnets 
in the vicinity of the saturation field $H_s$.
\cite{schulenburg,schnack,schmidt,mzh04,richter,1d,derzhko}  

At high fields Heisenberg antiferromagnets exhibit 
a phase transition between a fully polarized state and 
a state with reduced magnetization. A partially magnetized
state typically breaks spin-rotational symmetry about the
field direction and has a long-range order of transverse 
spin components. Since
the quantum ground state at $H>H_s$ is known exactly
one can also calculate
an exact spectrum of single magnon excitations $\omega({\bf k})$.
\cite{mattis} \ 
For ordinary nonfrustrated antiferromagnets
$\omega({\bf k})$ has a 
well defined minimum at a certain wave-vector $\bf Q$
with a gap $\Delta = H-H_s$. The high-field transition can 
be described as a Bose condensation of spin-flips 
with momenta ${\bf k}={\bf Q}$. \cite{batyev} \ 
The above simple picture fails, however, for frustrated magnets.
Reflecting degeneracy of classical ground states at $H<H_s$,
the magnon dispersion at $H>H_s$ has a continuous set
of degenerate minima. For two well-known frustrated models,
Heisenberg antiferromagnets on a face centered cubic lattice
and on a frustrated square lattice, single magnon spectra 
in the saturated phase are degenerate on lines in the Brillouin zone.
The phase transition at $H=H_s$ is, therefore, associated with
condensation of an infinite number of soft modes.
This yields an effective dimensionality reduction
and change in the critical behavior. \cite{jackeli} \ 
Geometrically frustrated 
antiferromagnets on kagom\'e, checkerboard, and pyrochlore
lattices exhibit even stronger degeneracy 
with a completely flat lowest
branch of magnons: $\omega({\bf k})\equiv H-H_s$. 
The theory of such phase transitions is discussed below on
two examples of the sawtooth chain and the kagom\'e antiferromagnet,
Fig.~1.

\section{Localized magnons}

\begin{figure}[t]
\begin{center}
\includegraphics[width=8cm]{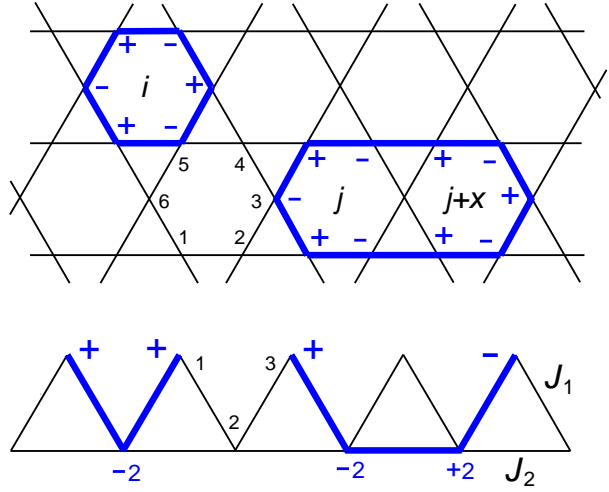}
\end{center}
\caption{
Kagom\'e lattice (top) and the sawtooth chain (bottom) with
localized magnons shown by thick lines.
Amplitudes and phases of spin-down states 
are indicated near each site. 
\label{geometry}} 
\end{figure}

We consider nearest-neighbor Heisenberg antiferromagnets in an 
external field
\begin{equation}
\hat{\cal H} = \sum_{\langle ij\rangle} J_{ij} 
{\bf S}_i\cdot {\bf S}_j
- {\bf H}\cdot \sum_i {\bf S}_i 
\label{Hamiltonian}
\end{equation}
with a general value $S$ of on-site spins. 
The exchange coupling constants $J_{ij}$ connect only 
the nearest neighbor sites. The couplings are all 
the same for a kagom\'e lattice $J_{ij}\equiv J$, while the 
sawtooth chain is described by two exchange constants: $J_1$ for 
base-vertex bonds and $J_2$ for base-base bonds, see Fig.~\ref{geometry}.
Above the saturation field $H_s$ the ground state of (\ref{Hamiltonian})
is a ferromagnetic vacuum $|0\rangle = |\uparrow\uparrow\uparrow...\rangle$,
where all spins are in a state with the maximum possible value
of $S_i^z =S$. The low-lying excitations are spin flips
$|i\rangle = S_i^-|0\rangle$. Diagonalizing the Heisenberg Hamiltonian 
(\ref{Hamiltonian}) within one-magnon subspace, which is possible
since $S^z_{\rm tot} = \sum_i S_i^z$ commutes with $\hat{\cal H}$,
one finds an exact one-particle spectrum. \cite{mattis}

The primitive unit cell of a kagom\'e 
lattice contains three sites. Accordingly, there exist three branches of
magnons for every wave-vector:
\begin{equation}
\omega_{1}({\bf k}) = H-6JS \ ,\ \ \ \omega_{2,3}({\bf k}) = H - 3JS \pm
JS\sqrt{3(1+2\gamma_{\bf k})} \ , 
\label{wkag}
\end{equation}
where $\gamma_{\bf k} = 1/6 \sum_l 
e^{i{\bf k}\cdot{\bf a}_l}$ is a sum over six nearest-neighbor 
sites on a triangular Bravais lattice. The lowest energy branch
$\omega_1({\bf k})$ has no dispersion and its energy vanishes 
at the saturation field $H_s=6JS$. Among the two dispersive branches 
the lower one $\omega_3({\bf k})$ has also a vanishing gap
at the saturation field.

In the case of the sawtooth chain, one, generally, finds
two dispersive branches of one-magnon excitations:
\begin{equation}
\omega_{1,2}(k) = H - 2J_1S - J_2S(1-\cos k) \pm 
S\sqrt{J_2^2(1-\cos k)^2+2J_1^2(1+\cos k)}\ .
\label{wsaw1}
\end{equation}
However, for a special choice of the coupling constants 
$J_2 = 0.5 J_1$ ($J_1\equiv J$), 
the lower branch becomes dispersionless:
\begin{equation}
\omega_1(k)=H-4JS\ ,\ \ \omega_2(k)=H - JS(1-\cos k) \ .
\label{wsaw}
\end{equation}
The above two branches are separated by a finite gap $\Delta\omega=2JS$.

The two formulated spin models are related to several 
magnetic materials. A kagom\'e lattice model is applicable, for example, 
for SrCr$_{9p}$Ga$_{12-9p}$O$_{19}$, \cite{scgo} while the sawtooth chain
describes magnetic delafossite
YCuO$_{2.5}$. \cite{delafossite} \ Unfortunately, both compounds have
rather large exchange constants $J\sim 500K$, which makes impossible
to reach their high-field regimes $H\sim H_s$.

The dispersionless excitations correspond in real space to localized states.
Spin flips are completely localized excitations in Ising models,
while for Heisenberg interaction they hop between adjacent lattice sites 
and generally acquire a finite dispersion.
In the two considered examples localization of spin flips is determined 
by lattice topology. For a kagom\'e lattice a simplest localized
state corresponds to  a magnon trapped on a hexagon void. Its wave 
function is given by
\begin{equation}
|\varphi_i\rangle =  \frac{1}{\sqrt{12S}} 
\sum_{n=1}^6 (-1)^{n-1}S^-_{ni}|0\rangle \ ,
\label{localK}
\end{equation}
where numbering of sites inside the $i^{\rm th}$ hexagon goes
counterclockwise starting with
the lower left corner, see Fig.~\ref{geometry}. 
An outside spin on one of the triangles
surrounding hexagon is connected to two sites, where a localized
magnon has equal amplitudes and opposite phases. This produces
destructive interference and the amplitude of spin flip
on surrounding sites vanishes.
The energy of the localized magnon (\ref{localK}) is
$H-6JS$, that is the same as $\omega_1({\bf k})$. 
More complicated eigenstates with the same energy
are constructed by taking closed 
loops with even number of spins, which contain only two sites 
of any crossed triangle. The wave-function of a 
localized state has equal amplitudes on all chosen
sites with alternating phases $0$ and $\pi$. 
An example of a spin flip localized on
two hexagons is shown in Fig.~\ref{geometry}. 
The wave-function of such a state
is a linear combination of simple hexagon states:
$|\varphi_{j,j+x}\rangle\propto (|\varphi_j\rangle + 
|\varphi_{j+x}\rangle)$.

The localized modes on smallest hexagons can be used as 
a nonorthogonal basis in the subspace of one-magnon states 
from the dispersionless branch (\ref{wkag}). 
To show this we note, first, that there are
as many hexagons $N_h=N/3$ on an $N$-site lattice as the number of states
in the Brillouin zone. Second, there is only one linear relation
$\sum_ia_i|\varphi_i\rangle =0$ between hexagon states for a 
cluster with periodic boundary conditions. It is constructed 
using the following arguments. An arbitrary site is 
shared between only two hexagons $i$ and $j$. Wave-functions
of the corresponding localized states have opposite signs 
for spin-flip amplitudes on the chosen site. Accordingly, the
two amplitudes $a_i$ and $a_j$ from a linear relation 
must be equal. This property, when extended by induction,
leads to a unique 
linear relation between hexagon states: 
a sum of all states with equal amplitudes $a_i={\rm const}$. 
The torus topology of a periodic cluster 
allows presence of a few other one-magnon states
with the same energy $H-6JS$, which cannot be decomposed into a 
hexagon basis.
Such localized magnon states correspond to closed
lines with nontrivial winding around the cluster.
The localized magnon loops obtained as linear combinations
of hexagon states
have, in contrast, zero winding and are contractable
into a point.
In the thermodynamic limit $N\rightarrow\infty$
presence of a few states with nontrivial winding is unimportant
and the localized hexagon states can be used as a real-space
basis for the lowest branch in Eq.~(\ref{wkag}). 

For the sawtooth chain a localized magnon is trapped in a valley between 
two triangles. Its wave-function is 
\begin{equation}
|\varphi_i\rangle  =  \frac{1}{\sqrt{12S}}
(S^-_{1i} - 2 S^-_{2i} +S^-_{3i}) |{\rm 0}\rangle \ .
\label{localS}
\end{equation}
The numbering of spins inside one valley is shown in Fig.~\ref{geometry}
with identity ${\bf S}_{1i}\equiv{\bf S}_{3,i-1}$.
The probability for a spin-flip to jump to
an outside site vanishes in 
the state (\ref{localS}) due to the relation
$J_1=2J_2$. The $N/2$ valley states again form a 
complete nonorthogonal basis
in the subspace of the lowest branch $\omega_1(k)$ (\ref{wsaw}).
Since bottom (base) spins are not shared between adjacent
valleys, the linear independence of states (\ref{localS})
is fulfilled even for finite periodic chains.
A more extended localized state on a sawtooth chain shown 
in Fig.~\ref{geometry} is 
a linear combination of states in the two adjacent valleys.

\begin{figure}[t]
\begin{center}
\includegraphics[width=9cm]{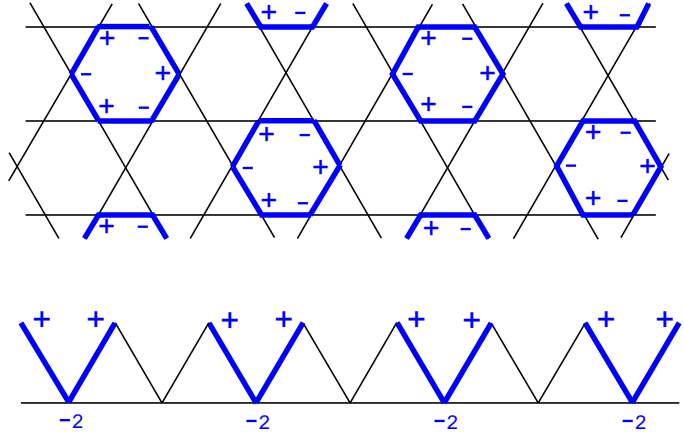}
\end{center}
\caption{
Magnon crystals built from localized spin flips
for a kagom\'e lattice (top) and the sawtooth chain (bottom).
\label{crystal}} 
\end{figure}

Localized one-magnon states allow to construct a class of exact
multiparticle states. Configurations, where
localized spin flips occupy isolated clusters, become
exact multimagnon states of the quantum spin Hamiltonian.
Note, that for nonfrustrated quantum magnets
in two or three dimensions, it is not possible
to construct exact multiparticle states beyond the
two-magnon subsector. \cite{mattis} \ 
Among all exact eigenstates with noninteracting localized magnons,
states with the highest density of 
spin flips play a special role. These are constructed 
as close-packed structures of 
magnons localized on smallest clusters: hexagons or valleys.
For a kagom\'e lattice the close-packed structure 
is constructed by putting localized
magnons on every third hexagon, see Fig.~\ref{crystal}.
Such a magnon crystal with $N_{\rm max}=N/9$ localized spin-flips 
breaks a three-fold translational symmetry of the magnetic Hamiltonian.
Transverse components of spin operators on sites
belonging to different 
localized magnons remain uncorrelated.
The magnon crystal does not break, therefore, spin-rotational 
symmetry about $z$-axis.
A magnon crystal for the sawtooth chain has $N_{\rm max}=N/4$ 
localized spin-flips and is two-fold degenerate. 

Since isolated localized 
magnons do not interact with each other, it is plausible 
that they correspond to the lowest energy states in every 
$n$-magnon subsector with $n\leq N_{\rm max}$. The proof of 
the above property is easily formulated for lattices with a uniform
exchange constant $J_{ij}=J$ like kagom\'e, checkerboard, and 
pyrochlore antiferromagnets. \cite{schnack,schmidt,mzh04} \ 
Let us consider, for simplicity, only $S=1/2$  case.
The Heisenberg Hamiltonian (\ref{Hamiltonian})
can be split into
an Ising and a transverse parts according to
\begin{equation}
\hat{\cal H}^{zz} = 
J \sum_{\langle ij\rangle} S_i^z S_j^z  - H\sum_i S_i^z \ , \ \ 
\hat{\cal H}^{\perp} = \frac{1}{2} J
\sum_{\langle ij\rangle} 
\bigl(S_i^+ S_j^- + S_i^-S_j^+\bigr) \ . 
\label{Hsplit1}
\end{equation}
Below we always subtract energy of the ferromagnetic state 
from $\hat{\cal H}^{zz}$.
A quantum state of $n$ isolated localized magnons  is 
an eigenstate of $\hat{\cal H}$ with $E_n = n(H-H_s)$,
where $H_s = 3J$ for a spin-1/2 antiferromagnet on a kagom\'e lattice.
In addition, such a state is also  an eigenstate 
of $\hat{\cal H}^{zz}$ and 
$\hat{\cal H}^\perp$ with $E_n^{zz} = n(H-2J)$ 
and $E_n^\perp = -nJ$, respectively. The idea of the proof is to show
that for an arbitrary $n$-magnon state: 
$\langle\hat{\cal H}^{zz}\rangle\geq E_n^{zz}$ and
$\langle\hat{\cal H}^\perp\rangle\geq E_n^\perp$.

For $\hat{\cal H}^{zz}$ the formulated relation holds trivially:
once all spin flips occupy different bonds the Ising
part of the Hamiltonian has the minimal energy:
$E_n^{zz}=n(H-\frac{1}{2}zJ)$, where $z=4$ is the number of nearest
neighbors for a kagom\'e lattice.
In order to show that a similar inequality holds
for the transverse part $\hat{\cal H}^\perp$ 
we map the $n$-magnon subspace of a spin-1/2 model onto the Hilbert space
of $n$ hard-core bosons ${\cal B}_n$. The transverse part 
of the Heisenberg Hamiltonian is, then, the kinetic 
energy of bosons $\hat{\cal H}^\perp\equiv \hat{\cal K}$. 
In addition, we define the Hilbert space of $n$ bosons 
without the hard-core constraint ${\cal B}_{n0}$
with an evident relation
${\cal B}_n\subset{\cal B}_{n0}$.
The minimum of the kinetic energy is easily found in 
${\cal B}_{n0}$: all $n$ particles must occupy one of 
the lowest energy single-particle states with the kinetic energy
$E_{\cal K} = -J$ found from Eq.~(\ref{wkag}).
If the hard-core constraint is imposed, 
then, from the variational principle, the expectation
value of $\hat{\cal K}$ can only increase: 
$\min\langle \hat{\cal K} \rangle_{{\cal B}_n}  \geq 
\min\langle \hat{\cal K} \rangle_{{\cal B}_{n0}} =-nJ=E_n^\perp$.

The above proof does not immediately apply to the sawtooth chain,
which has nonequivalent bonds. 
A localized magnon (\ref{localS}) is an 
eigenstate of the total Hamiltonian
$\hat{\cal H}$, but not of  
$\hat{\cal H}^\perp$ or $\hat{\cal H}^{zz}$. 
In order to solve this problem we rewrite the Ising part in 
Eq.~(\ref{Hsplit1}) and split the Heisenberg Hamiltonian 
as $\hat{\cal H} = \hat{\cal H}_1 + \hat{\cal H}_2$, 
\begin{eqnarray}
&& \hat{\cal H}_1 = 
\sum_i \varepsilon_i \Bigl(\frac{1}{2}-S^z_i\Bigr)+\hat{\cal H}^\perp \ , \
\ \ \ \varepsilon_i = H - \frac{1}{2} \sum_j J_{ij} \ ,
\nonumber \\
&& \hat{\cal H}_2 = 
\sum_{\langle ij\rangle} J_{ij} \Bigl(\frac{1}{2}-S^z_i\Bigr) 
\Bigl(\frac{1}{2}-S^z_j\Bigr) \ .
\label{Hsplit2}
\end{eqnarray}
Localized states (\ref{localS}) are eigenstates of 
$\hat{\cal H}_1$ with the eigenvalue $H-2J$ ($S=1/2$) and, trivially, 
of $\hat{\cal H}_2$ with zero eigenvalue.  
Then, the proof for the sawtooth chain copies the above
arguments for the kagom\'e antiferromagnet applied
to $\hat{\cal H}_1$ and $\hat{\cal H}_2$ instead of
$\hat{\cal H}^\perp$ and $\hat{\cal H}^{zz}$. 
Similar conclusion has been earlier obtained using a somewhat different approach 
in Ref.~\citen{schmidt}.

After establishing that isolated localized magnons are the lowest
energy states, the high-field magnetization process of 
geometrically frustrated spin systems
at $T=0$ can be understood as follows.
Above $H_s$ all magnetization
subsectors are separated from the ferromagnetic ground
state by finite gaps $E_n=n(H-H_s)$. At the saturation field
all gaps in subsectors with $n\leq N_{\rm max}$ vanish
and a huge degeneracy of the quantum ground state develops
for frustrated antiferromagnets.
Below $H_s$ the subsector with the largest possible density of
localized magnons $n=N_{\rm max}$ corresponds to the lowest energy.
Therefore, there is a finite jump of the magnetization at
$H=H_s$ between a saturated phase and a magnon crystal state.
Such universal jumps with $\Delta M = N_{\rm max}/N$
have been found in numerical exact diagonalization 
studies of finite clusters of the sawtooth
chain, kagom\'e and checkerboard lattice antiferromagnets and
several other frustrated one-dimensional models.
\cite{schulenburg,schnack,richter,1d,derzhko} \  It is 
intuitively clear that noninteracting localized magnons
being the lowest energy state in the corresponding magnetization
subsectors make also
the largest contribution to the partition function
at low temperatures. Moreover, 
since the localized excitations
have no dynamics their statistics is equivalent to statistics
of specially chosen classical particles. Using such a mapping one
can find an exact low-temperature behavior of quantum 
frustrated magnets in the vicinity of the saturation field. \cite{mzh04} \ 
Corrections to such an asymptotic behavior
should be exponentially small once the isolated localized
magnons are separated from 
from higher-energy propagating states by finite gaps.
Below we consider separately
the low-temperature properties of
the sawtooth chain and kagom\'e antiferromagnet.

\section{The sawtooth chain}

Contribution of degenerate levels to the partition
function is determined
by dimensionality of the corresponding subspace of the total
Hilbert space of a quantum system.
In order to calculate the dimensionality we can utilize an
arbitrary basis for degenerate states. Such a basis has to be complete
but not necessarily orthogonal.
Among different configurations of noninteracting
localized magnons only states
formed by magnons localized in isolated valleys have to be counted
for the sawtooth chain.
Extended localized states, such as a
state on the right-hand-side of Fig.~\ref{geometry},
are represented as linear combinations of the basis valley
states.
The multimagnon `valley' states can be associated with dimer coverings 
of a base chain: once a localized spin-flip (\ref{localS})
occupies a given valley
or a site of the base chain it is impossible to put
another localized magnon to the same or the adjacent valleys.
The hard-dimer states in a given magnetization subsector
are linearly independent.
This property can be proven by induction using linear independence of
the valley states (\ref{localS}) in the one-magnon subsector.
In the following, we estimate gaps which separate
the hard-dimer states from  a
continuum of the scattering states and then 
calculate the low-temperature thermodynamics of  
the sawtooth chain. 

\subsection{Two-magnon bound states}

The scattering states of two magnons can be in principle
calculated exactly using a general method discussed
in Ref.~\citen{mattis}. 
We shall adopt instead a variational approach, which is much simpler
for the present problem.
The two branches of single magnon excitations for the sawtooth chain 
Eq.~(\ref{wsaw}) are separated by a finite gap $\Delta\omega=2JS$. 
The low-energy scattering states are, therefore,
formed by magnons from the lowest branch $\omega_1(k)$.
In the two-magnon sector there are 
$\frac{1}{2}(N/2)(N/2+1)$ states constructed from
$N/2$ one-magnon states of the dispersionless branch.
The number of the basis hard-dimer states is
$\frac{1}{2}(N/2)(N/2-3)$. The difference between the two yields the number
of low-energy scattering states, which is equal to $N$. In real space 
representation the above scattering states correspond to
localized magnons occupying either the same $|\varphi_i^2\rangle$
or the adjacent $|\varphi_i\varphi_{i+1}\rangle$ valleys.
We again assume $S=1/2$ for simplicity and construct accordingly 
the following variational basis
for the low-energy interacting two-particle states: 
\begin{eqnarray}
&& |\psi_i\rangle = \frac{1}{\sqrt{35}} 
(S_{1i}^- - 2S_{2i}^- + S_{3i}^-)
(S_{1,i+1}^- - 2S_{2,i+1}^- + S_{3,i+1}^-)|0\rangle 
\simeq  |\varphi_i\varphi_{i+1}\rangle 
\nonumber \\
&& |\widetilde{\psi}_i\rangle = \frac{1}{6} 
(S_{1i}^- - 2S_{2i}^- + S_{3i}^-)^2 |0\rangle 
 \simeq |\varphi_i^2\rangle \ .
\end{eqnarray}
The exclusion principle of spin-flips has been used for normalization
in the above equation.
States $|\psi_i\rangle$ and $|\widetilde{\psi}_i\rangle$ have 
nonvanishing overlaps:
\begin{equation}
\langle\psi_i|\psi_{i\pm 1}\rangle = 1/35 \ , \ \ \ 
\langle\psi_i|\widetilde{\psi}_i\rangle = 
\langle\psi_i|\widetilde{\psi}_{i+1}\rangle = 
\sqrt{5/63} \ .
\end{equation}
The nonzero matrix elements of the Hamiltonian between
these states are
\begin{eqnarray}
&& \langle\psi_i|\hat{\cal H}|\psi_i\rangle = 2H-\frac{24}{7}J \ ,
\ \ \langle\widetilde{\psi}_i|\hat{\cal H}|\widetilde{\psi}_i\rangle = 
2H-2J \ , \ \ 
\langle\psi_{i\pm 1}|\hat{\cal H}|\psi_i\rangle = \frac{2H-4J}{\sqrt{35}}\ ,
\nonumber \\
&& 
\langle\widetilde{\psi}_{i\pm 1}|\hat{\cal H}|\widetilde{\psi}_i\rangle 
= \frac{1}{9}J \ , \ \ 
\langle\widetilde{\psi}_i|\hat{\cal H}|\psi_i\rangle = 
\langle\widetilde{\psi}_{i+1}|\hat{\cal H}|\psi_i\rangle = 
\frac{10H-14J}{3\sqrt{35}} \ .
\label{matrixEs}
\end{eqnarray}
A simple variational ansatz for a propagating state of two
localized magnons in adjacent valleys is constructed as 
$|k\rangle = \sum_i e^{-ikr_i}|\psi_i\rangle$. It has the 
following energy 
\begin{equation}
E(k)=\frac{\langle k|\hat{\cal H}|k\rangle}{\langle k|k\rangle}
= 2(H-2J) + \frac{20J}{35+2\cos k} \ .
\label{2mag}
\end{equation}
Two interacting localized magnons cannot scatter on each other
because of an absence of propagating one-particle states at low energies.
Instead they form a bound two-magnon pair with energy $E(k)$. 
Finite dispersion of such pairs comes from assisted hopping
of adjacent localized magnons.
The minimal value of $E(k)$ is reached at $k=0$ with the gap
$\Delta=20/37J\approx 0.54J$ separating the low-energy 
boundary $E_2=2(H-2J)$ from the higher energy states.
Numerical minimization
of $E(k)$ for an improved variational ansatz $|k\rangle=
\sum_i e^{-ikr_i}(|\psi_i\rangle+c|\widetilde{\psi}_i\rangle)$ 
yields $\Delta \approx 0.44J$ 
with an optimal value $c\approx -0.24$.
This value of the gap compares very well with 
the numerical diagonalization result $0.42J$. \cite{1d} \ 
The gap is determined mostly by a nearest-neighbor repulsion of 
localized magnons in adjacent valleys. The dispersion of bound 
two-magnon complexes $E(k)$ is weak and does not exceed 7\%.
We conjecture that the intervalley repulsion leads to a similar 
behavior in all $n$-magnon sectors: the lowest 
energy states with $E_n=n(H-2J)$ are separated from the 
scattering states by a finite gap $\Delta_n=O(J)$. At 
temperatures $T\ll\min_n\{\Delta_n\}$ one can neglect the higher energy 
states and consider only the contribution of isolated localized 
magnons. The latter problem is equivalent to a one-dimensional
lattice gas of particles with energies $H-2J$ and 
on-site and nearest-neighbor exclusion principle, which is also
known as a classical 
hard-dimer model.

\subsection{Low-temperature behavior}

We consider classical  hard  dimers on a periodic chain of length $L=N/2$,
$N$ being the number of spins in the sawtooth chain.
The partition function of this model is
\begin{equation}
{\cal Z} = \sum_{\{\sigma\}} 
\exp\left[\frac{\mu}{T}\sum_i\sigma_i\right]
\prod_{\langle ij\rangle} (1-\sigma_i\sigma_j) \ ,
\label{Zsaw}
\end{equation}
where $\mu=H_s-H$ is the chemical potential and 
$\sigma_i=0,1$ are the occupation numbers for dimers and $k_B\equiv 1$.
The nearest-neighbor exclusion principle is imposed by 
the last term. The partition function (\ref{Zsaw}) 
can be exactly calculated using the standard transfer matrix approach.
In the thermodynamic limit the free energy is
determined by the largest eigenvalue of the transfer matrix:
\begin{equation}
{\cal F}/N = -\frac{1}{2}\,T
\ln\biggl(\frac{1}{2} + \sqrt{\frac{1}{4} + e^{\mu/T}}
\biggr) .
\label{Fsaw}
\end{equation}
The entropy ${\cal S}= -\partial{\cal F}/\partial T$
depends on external magnetic field and temperature via $\mu/T$:
\begin{equation}
{\cal S}/N = \frac{1}{2}
\ln\Bigl(1 + \sqrt{1 + 4e^{\mu/T}}\Bigr) - \frac{1}{2}\ln 2
- \frac{\mu}{4T}\biggl(1 - \frac{1}{\sqrt{1+4e^{\mu/T}}} 
\biggr) .
\label{Ssaw}
\end{equation}
During an adiabatic process the combination
$(H-H_s)/T$ remains constant and, consequently,
temperature drops to zero as $H\to H_s$. 
Such a very strong magnetocaloric effect is a direct consequence
of the condensation of a macroscopic number of soft modes 
at $H=H_s$. \cite{mzh03} \ At $H=H_s$ the entropy 
remains finite down to $T=0$ and has a universal value:
\begin{equation}
{\cal S}/N=\frac{1}{2}\ln\biggl(\frac{1+\sqrt{5}}{2}\biggr)\approx 
0.34712\,\ln 2 \ . 
\label{ST0}
\end{equation}
The localized magnons of the sawtooth chain 
behave, therefore, similar to paramagnetic degrees of freedom in
an effective field $h=H-H_s$.
The major difference with an ideal paramagnet is that 
the entropy of the latter
at $h=0$ depends on a spin length,
whereas  the residual entropy of frustrated magnets at $H=H_s$
has a geometric origin. 

\begin{figure}[t]
\begin{center}
\includegraphics[width=9cm]{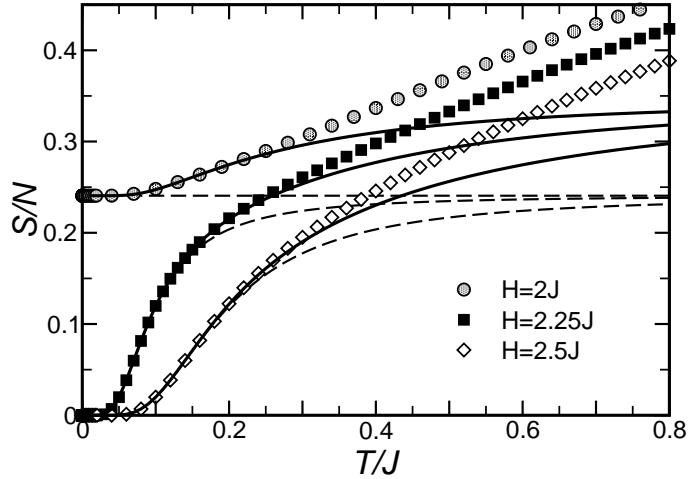}
\end{center}
\caption{
Temperature dependence of the entropy of the sawtooth chain
in magnetic field. Symbols are exact diagonalization data,\cite{1d}
dashed lines correspond to the hard-dimer model, solid lines are
obtained in the effective Ising model.
\label{sawTscanS}} 
\end{figure}

Using the transfer matrix technique, one can also 
calculate the dimer-dimer correlation function.  
The result is 
\begin{equation}
  \langle \sigma_i \sigma_j \rangle = \bar{\sigma}^2
  + (-1)^{i-j} \bar{\sigma} (1-\bar{\sigma}) 
  e^{-|r_i - r_j | / \xi}, 
\end{equation}
where the average dimer density $\bar{\sigma}$ and 
the correlation length $\xi$ are given by 
\begin{eqnarray}
  \bar{\sigma} &=&
  \sin^2 \frac{\gamma}{2}
  = \frac{2 e^{\mu/T}}{1+4e^{\mu/T}+\sqrt{1+4e^{\mu/T}}} , 
  \qquad \tan \gamma \equiv 2e^{\mu/2T} \nonumber \\
  \frac{1}{\xi} &=&
  2 \log \left( \cot \frac{\gamma}{2} \right) 
  = \log \left( \frac{\sqrt{1+4e^{\mu/T}}+1}{\sqrt{1+4e^{\mu/T}}-1} \right) . 
\end{eqnarray}
The correlation length exponentially diverges $\xi \sim e^{\mu/2T}$ as
$T\to 0$.

The specific heat is obtained by taking temperature derivative 
of the internal energy
${\cal E} = -1/2\,\mu \bar{\sigma}$:
\begin{equation}
 C/N = \frac{\mu^2}{8 T^2} \sin^2 \gamma \cos \gamma 
 = \frac{\mu^2}{2 T^2}
 \frac{e^{\mu/T}}  
 {(1+4e^{\mu/T})^{3/2}}\  . 
\end{equation}
The specific heat as a function of $x=\mu/T$
has two peaks at $x=x_\pm$ determined by solutions of 
the following equation:
\begin{equation}
x+2=2e^x(x-4) \ \ \ \ {\rm for}\ \  
x_+=4.0526\ ,\  x_-=-2.8159 \ .
\label{saw_peaks}
\end{equation}

\begin{figure}[t]
\begin{center}
\includegraphics[width=9cm]{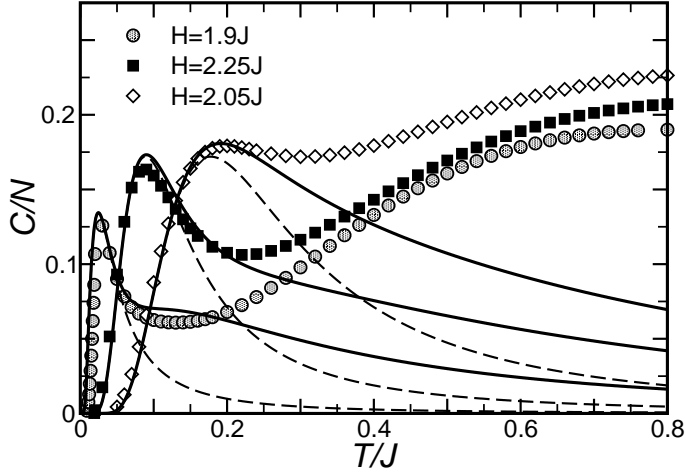}
\end{center}
\caption{
Temperature dependence of the specific heat of the sawtooth chain
in magnetic field. Symbols are exact diagonalization data,\cite{1d}
dashed lines correspond to the hard-dimer model, solid lines are
obtained in the effective Ising model.
\label{sawTscanC}} 
\end{figure}

\begin{figure}[t]
\begin{center}
\includegraphics[width=9cm]{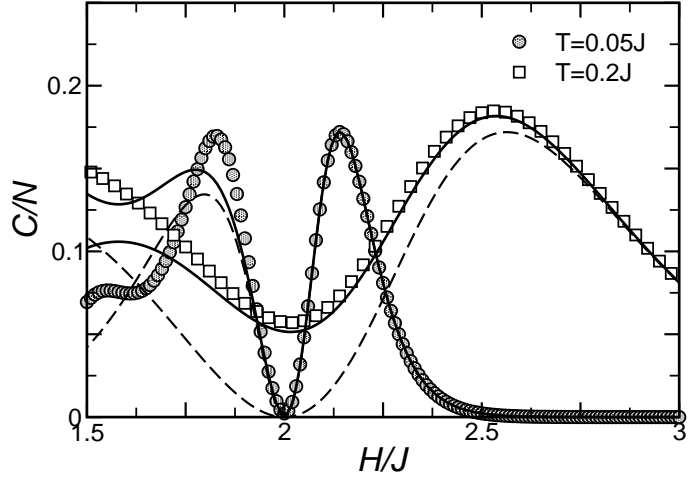}
\end{center}
\caption{
Field dependence of the specific heat of the sawtooth chain.
Symbols are numerical data,\cite{1d}
dashed lines correspond to the hard-dimer model, full curves are
calculated in the effective Ising model.
\label{sawHscanC}} 
\end{figure}

Comparison of the entropy and the specific heat of the sawtooth chain
calculated in the hard-dimer model 
and by a full numerical diagonalization 
of a cluster with $N=20$ sites \cite{1d}
is shown in Figs.~\ref{sawTscanS}--\ref{sawHscanC}.
The analytic and the numerical results agree well with each other below $T^*\sim 0.1J$.
In particular, the hard-dimer representation correctly 
reproduces the residual entropy at $H=H_s$ (Fig.~\ref{sawTscanS}).
At  $T>0.1J$ a thermal contribution of the higher energy states becomes
significant.
Those states can be partially taken into account by using analogy with
Ising models near the saturation field, which are also mapped
to statistical models of hard-core objects. \cite{metcalf,baxter} \ 
For the sawtooth chain such a generalized Ising representation
can be constructed because of a weak dispersion of bound 
two magnon pairs (\ref{2mag}). 
Once the $k$-dependence of $E(k)$ is ignored, there are no
zero-point fluctuations and we end up again with a model
of classical hard-core particles, where the nearest-neighbor exclusion is
replaced by an Ising type repulsion:
\begin{equation}
{\cal H} = \sum_i \bigl(-\mu\sigma_i + V \sigma_i\sigma_{i+1}\bigr) \ ,
\ \ \ \ \sigma_i=0,1 \ .
\label{Hclass}
\end{equation}
For the strength of the Ising interaction we have to choose
the gap, which separates bound two-magnon pairs from
noninteracting localized magnons: $V=\Delta\approx 0.44J$.
The classical Ising model (\ref{Hclass}) is solved  exactly
by the transfer-matrix method with the following result for 
the free energy: 
\begin{equation}
{\cal F}/N = -\frac{1}{2}\,T
\ln\biggl(\frac{1}{2}\bigl[1+e^{-(V-\mu)/T}\bigr] + \sqrt{\frac{1}{4}
\bigl[1-e^{-(V-\mu)/T}\bigr]^2  + e^{\mu/T}}
\biggr) .
\label{Fsaw2}
\end{equation}
The above expression transforms into the hard-dimer result (\ref{Fsaw}) 
for $T\ll V$. 
Comparison between numerical exact diagonalization results
and the two analytic approximations is presented in Fig.~\ref{sawTscanS}.
The entropy calculated in the Ising model shows 
a nice agreement
with the numerical data up to a remarkably high $T^*\sim 0.3J$.
Note, that the Ising repulsion $V$ is not a fitting
parameter, but has been calculated above.

The specific heat of the sawtooth chain
as a function of temperature in a constant magnetic field 
$H\sim H_s$ (Fig.~\ref{sawTscanC})
has a peak at $T_m = (H_s-H)/x_\pm$, where $\pm$ correspond to 
$H<H_s$ and $H>H_s$, respectively. Such a low-temperature peak 
appears due to a freezing of localized degrees 
of freedom and is well described
by both analytic approaches. At higher temperatures $T\sim J$, 
the specific heat has a second peak, which reflects development
of one-dimensional short-range spin correlations.  Its presence is
not captured, of course, by the low-temperature hard-dimer mapping.
Experimental observation of the double-peak structure in 
temperature dependence of the specific heat of real frustrated
magnetic compounds can be a clear
sign of localized magnons.

\begin{figure}[t]
\begin{center}
\includegraphics[width=9cm]{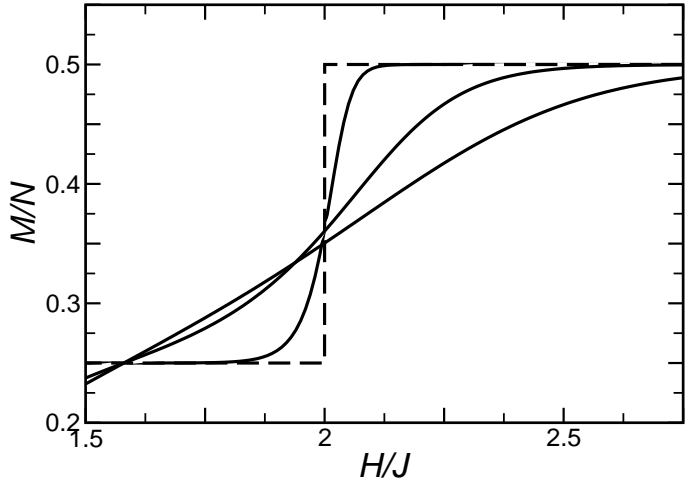}
\end{center}
\caption{
Magnetization curves of the sawtooth chain
obtained in the effective Ising model. Full lines correspond from
top to bottom to $T=0.02J$, $0.05J$, $0.1J$, and $0.2J$;
dashed line is the zero-temperature magnetization jump. 
\label{sawHscanM}} 
\end{figure}

Field dependence of the specific heat is presented 
in Fig.~\ref{sawHscanC}.  
$C(H)$ has a deep minimum at the saturation field.
The minimum occurs due to a weak temperature dependence of
${\cal S}(T,H_s)$, which is completely $T$-independent in
the hard-dimer model. According to Eq.~(\ref{saw_peaks}), 
the high- and low-field peaks in the specific heat correspond to
$H_m = H_s -Tx_\pm$. Comparison between the three types of 
the results for the specific heat shown in Fig.~\ref{sawHscanC} 
allows also to determine the field range, where the localized magnons
play a dominant role: $H>0.8H_s$.

The zero-temperature jump between a half-magnetization
plateau and the fully saturated phase
is smeared at  finite temperatures.
According to the hard-dimer representation, 
the magnetization curves $M(H,T)= (1-\bar{\sigma})/2$ cross
at $H=H_s$ with the universal value
$M(T,H_s)/N = (1+\sqrt{5})/4\sqrt{5}$.
The finite-temperature width of the jump is 
approximately $\Delta H\sim 4T$.
The above value of the magnetization 
corresponds to an average density $n=(\sqrt{5}-1)/4\sqrt{5}
\approx 0.1382$ of localized magnons.
Magnetization curves of the sawtooth chain 
are illustrated in Fig.~\ref{sawHscanM} by using
the effective Ising model, which give 
more accurate results for the selected temperatures than the
hard-dimer mapping.

\section{Kagom\'e antiferromagnet}

Following our analysis of the sawtooth chain,
we define a class of hard-hexagon states
for a kagom\'e antiferromagnet near the saturation.
They are built from magnons localized on
smallest hexagon voids of a kagom\'e lattice, such that no two hexagons
have common sites. 
The hard-hexagon states are linearly independent in the thermodynamic limit
with at most
one linear relation in every magnetization subsector.
The centers of hexagons form a triangular lattice and the above 
states are, therefore, mapped onto a classical gas of hard-core particles
on a triangular lattice with the nearest-neighbor exclusion 
principle. The latter is a famous exactly solvable model \cite{baxter}
and, therefore, a great deal of the exact information
is known about such states. However, in order
to apply the available results to a kagom\'e antiferromagnet 
we first have to show that 
(i) the hard-hexagon states are separated
by finite gaps from the higher-energy states
and (ii) there are no other low-energy states in every
magnetization subsector with $n\leq N_{\rm max}$ or their
contribution is vanishingly small.

\subsection{Two-magnon bound states}

Low energy part of the two-magnon subsector 
of a kagom\'e antiferromagnet consists of 
$\frac{1}{2}(N/3)(N/3+1)$ states constructed from one-particle 
states of the flat branch $\omega_1({\bf k})$. These include
$\frac{1}{2}(N/3)(N/3-7)$ hard-hexagon states. 
Additional $4N/3$ interacting or scattering states
correspond to $N$ states of localized magnons, which occupy 
adjacent hexagons, $|\varphi_i\varphi_{i+a_l}\rangle$,  
and to $N/3$ states with two spin flips on the same hexagon, 
$|\varphi_i^2\rangle$. In a simple treatment of a
spin-1/2 model we neglect
the latter states and define the following 
real-space basis for the former $N$ states:
\begin{equation}
|\psi_{li}\rangle  =  \frac{1}{\sqrt{35}} 
\sum_{n,n'=1}^6 (-1)^{n+n'} S^-_{ni} S^-_{n'i+a_l} 
|0\rangle \ ,  
\label{2magK}
\end{equation}
where $l=1$--3 and ${\bf a}_1=(1,0)$ and 
${\bf a}_{2,3}=(\pm 1/2,\sqrt{3}/2)$
are three nearest-neighbor sites on a triangular lattice
of hexagons. The normalization factor in 
Eq.~(\ref{2magK}) takes into account the exclusion
principle of spin flips. The overlap matrix elements 
for the above states are
\begin{equation}
\langle\psi_{2i}|\psi_{1i}\rangle = -1/7\ , \ \ \ 
\langle\psi_{3i}|\psi_{1i}\rangle = 
\langle\psi_{li\pm a_{l'}}|\psi_{li}\rangle =  1/35 \ .
\label{overlapK}
\end{equation}
The Hamiltonian (\ref{Hamiltonian}) has the following
nonzero matrix elements
\begin{equation}
\langle\psi_{1i}|\hat{\cal H}|\psi_{1i}\rangle = 2H -\frac{40}{7}J, \ \
\langle\psi_{2i}|\hat{\cal H}|\psi_{1i}\rangle = -\frac{
10H -29J}{35}, \ \ 
\langle\psi_{3i}|\hat{\cal H}|\psi_{1i}\rangle = \frac{
2H -6J}{35}.
\label{matrixK}
\end{equation}
All other nonvanishing matrix elements are obtained 
from Eqs.~(\ref{overlapK})
and (\ref{matrixK}) by applying symmetry operations of a kagom\'e lattice.
Using ansatz 
$|{\bf k}\rangle = \sum_{il} e^{-i{\bf k}{\bf r}_i} 
c_l |\psi_{li}\rangle$ 
for a bound state of two magnons we calculate its energy as
\begin{eqnarray}
&& E({\bf k}) = \frac{\langle{\bf k}|\hat{\cal H}|{\bf k}\rangle}
{\langle{\bf k}|{\bf k}\rangle} = 2(H-3J) + \varepsilon({\bf k}) \ ,
\\ 
&& \varepsilon({\bf k}) =
\frac{2}{7}\;\frac{|c_1|^2 + |c_2|^2 + |c_3|^2 - \frac{1}{10} P_{\bf k}}
{(|c_1|^2 + |c_2|^2 + |c_3|^2)[1+\frac{2}{35}
(\cos k_1+\cos k_2+\cos k_3)]-\frac{1}{7}P_{\bf k}
+\frac{1}{35}Q_{\bf k}} \ , 
\nonumber \\[0.5mm] 
&& P_{\bf k} = c_1c^*_2(1+e^{-ik_3}) + c_2c^*_3(1+e^{ik_1}) +
c_3c^*_1(e^{-ik_1}+e^{ik_3}) + {\rm c.\,c.} \ ,
\nonumber \\ 
&& Q_{\bf k} = c_1c^*_2(e^{ik_1}+e^{-ik_2}) 
+ c_2c^*_3 (e^{ik_2}+e^{-ik_3}) 
+ c_3c^*_1(1+ e^{i(k_3-k_1)}) + {\rm c.\,c.} \ ,
\nonumber 
\label{2magEK}
\end{eqnarray}
where $k_i = {\bf k}\cdot {\bf a}_i$.
The minimum of $\varepsilon({\bf k})$ occurs at zero total momentum
${\bf k}=0$ for $c_1=c_2=c_3$ with a gap
$\Delta=0.24J$.
This value of the gap can be again used as 
an estimate for the repulsion between localized magnons
on adjacent hexagons, though dispersion of bound pairs $\sim 12$\%
is somewhat larger in two dimensions.

\begin{figure}[t]
\begin{center}
\includegraphics[width=9cm]{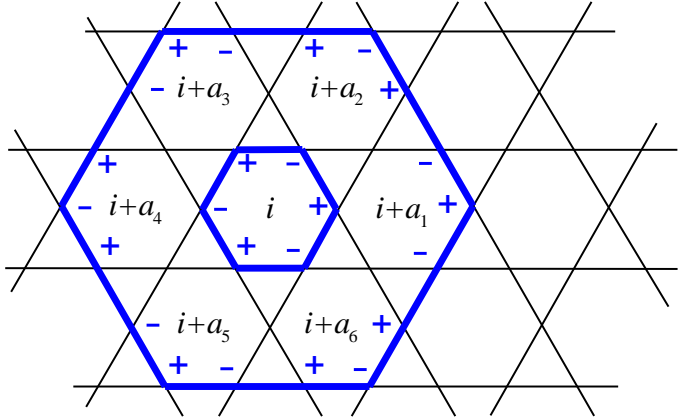}
\end{center}
\caption{
The defect state of two localized magnons, which breaks the hard-hexagon rule.
\label{defect}} 
\end{figure}

Among various states of $n$ isolated localized magnons
there are also states,
which do not obey the hard-hexagon rules.
The simplest example of such a `defect' state is shown in 
Fig.~\ref{defect}:
two localized magnons reside on a small $|\varphi_i\rangle$ and 
a large $|\varphi_L\rangle$ hexagons.
The large hexagon state is decomposed
into a basis hexagon states (\ref{localK}) as
\begin{equation}
|\varphi_L\rangle = \sum_{l=1}^6 |\varphi_{i+a_l}\rangle  
+ |\varphi_i\rangle \ .
\end{equation}
The two-particle state 
$|\varphi_L\varphi_i\rangle$ 
corresponds, therefore, to a special linear combination of states
(\ref{2magK}) and state $|\varphi_i^2\rangle$ and, obviously, breaks
the hard-hexagon rules.
By analogy with Fig.~\ref{defect} a general defect two-magnon state
$|\varphi_L\varphi_{L'}\rangle$ 
is constructed by drawing
two closed lines $L$ and $L'$ on a kagom\'e lattice
such that (i) the lines can host
localized magnons and (ii) one of the lines $L'$ lies inside
the area $A_L$ enclosed by the second line: $L'\subset A_L$.
Such states are, however, topologically equivalent to the two-magnon
state in Fig.~\ref{defect} and can be represented
as linear combinations of normal hard-hexagon states and 
the primitive defect state. In addition, if one or both lines
$L$ and $L'$ have nonzero winding around the cluster,
the corresponding two-magnon state cannot again be decomposed 
into hard-hexagon states (see discussion in Sec.~2).
In the two-magnon subsector the total number
of states breaking the hard-hexagon rule is $\propto N$, that is
a factor of $1/N$ smaller than the number of hard-hexagon states.
Similar estimate holds for the three-magnon subsector:
$O(N^2)$ defect states versus $O(N^3)$ hard-hexagon states
and so on. In addition, in magnetization subsectors with finite
densities of magnons the number of defect states is further reduced.
Indeed, the defect state shown in Fig.~\ref{defect} exists only
if there are no localized magnons on any of 12 small hexagons 
surrounding the larger loop. As the total density $n/N$ of (localized) 
magnons grows, there is less and less space for the composite
defect states. Finally, if there are more than
two magnons on every 19 hexagons or $n/N>2/(19\cdot 3)\approx 0.035$,
the number of defect states becomes significantly suppressed
and eventually goes to zero.

Another complication for the kagom\'e antiferromagnet comes
from the fact that 
the gap for one of the dispersive branches
in Eq.~(\ref{wkag}) also vanishes at the saturation field. The 
corresponding propagating magnon has the same energy
$\omega_3(0)=H-3J$ as localized magnons from 
the flat branch. The lowest-energy states in the two-magnon 
subsector contain apart from the isolated localized magnons
also superposition states of one localized magnon and
one propagating magnon. Such states form a continuum above
the low-energy threshold $E_2=2(H-3J)$.
The same is true for all $n$-magnon subsectors 
with $n\ll N$. Once $n$ becomes a finite fraction of $N$
the above picture changes. The low-energy propagating  
magnons experience  multiple scattering from an infinite 
number of localized magnons. Such scattering produces  
a finite shift of the energy of propagating magnons. 
An exact value of 
energy shift for $\omega_3({\bf k})$ depends on
a precise pattern of localized magnons. An explicit 
estimate can be obtained for an average `uniform' state 
of $n$-localized magnons, 
corresponding to a low-temperature ensemble of states  
with different translational patterns. We bosonize the 
spin Hamiltonian using, {\it e.g.\/}, the Dyson-Maleev
transformation: $S_i^+ = \sqrt{2S}(1-b_i^\dagger b_i/2S) b_i$,
$S_i^- = \sqrt{2S}b_i^\dagger$, for an arbitrary value
of on-site spin $S$.
Diagonalization of quadratic terms gives the excitation
spectrum (\ref{wkag}). 
Effect of interaction between magnons is treated in the lowest order
by using the Hartree-Fock approximation.
We define two averages: for an on-site density $m_1=\langle 
b_i^\dagger b_i\rangle$ and for a nearest-neighbor hopping 
$m_2=\langle b_i^\dagger b_j\rangle$.
After a mean-field 
decoupling of four-boson terms the effect of
interaction is reduced to a renormalization
$S\to (S-m_1+m_2)$ in the quadratic terms.
The lowest point of the dispersive branch is at 
$\omega_3(0) = H-6J(S-m_1+m_2)$, which reduces at 
the saturation field $H_s=6JS$ to $\Delta_d = 6J(m_1-m_2)$. 
For a single localized magnon (\ref{localK})
one finds $m_1=-m_2=1/6$.
Hence, the dispersive mode propagating through 
an `averaged ensemble' of $n$ localized magnons acquires 
a finite gap
$\Delta_d = 3Jn/N$, which separates the dispersive branch
from the localized magnons.
At low enough temperatures $T\ll\Delta_d$ the dispersive
modes have a negligible contribution to the thermodynamics of a kagom\'e
antiferromagnet.
Note, that characteristic temperature, where the above
approximation becomes valid, depends on an average density
of magnons and, therefore, on an applied magnetic field.

\subsection{Low-temperature behavior}

In the above analysis we have established that the hard-hexagon
description of lowest energy states is valid at sufficiently 
high densities of magnons. 
High or low means here in comparison to the
density of the magnon crystal: $n/N=1/9$.
Let us now discuss the thermodynamic properties of a spin-1/2
kagom\'e antiferromagnet in the hard-hexagon approximation.
The partition function of the hard-hexagon lattice gas is given by 
the same expression (\ref{Zsaw}), where the site index runs 
over ${\cal N}=N/3$ sites of a triangular lattice 
formed by centers of hexagon voids of the original kagom\'e 
lattice. This model was solved by Baxter with the help of 
a corner transfer matrix method. \cite{baxter} \ Specifically, 
the normalized partition function  ${\cal Z}^{1/{\cal N}}$ 
and the fugacity $z=e^{\mu/T}$ are expressed 
as functions of a real auxiliary parameter $x$:
at the high-field region $0<z<z_c$ ($-1<x<0$), 
\begin{eqnarray}
&& z = -x \frac{H^5(x)}{G^5(x)}\ , \ \ \  H(x) = \prod_{n=1}^\infty
[(1-x^{5n-4})(1-x^{5n-1})]^{-1}, \\[1mm]
&& G(x) = \prod_{n=1}^\infty [(1-x^{5n-3})(1-x^{5n-2})]^{-1}, \ \ \ 
Q(x) = \prod_{n=1}^\infty (1-x^n) \ ,
\nonumber \\
&& {\cal Z}^{1/{\cal N}}= \frac{H^3(x)Q^2(x^5)}{G^2(x)}\prod_{n=1}^\infty
\frac{(1-x^{6n-4})(1-x^{6n-3})^2
(1-x^{6n-2})}{(1-x^{6n-5})(1-x^{6n-1})(1-x^{6n})^2} \ ;
\nonumber 
\end{eqnarray}
whereas at the low-field region $z>z_c$ ($0<x<1$), 
\begin{eqnarray}
&& z = x^{-1} \frac{G^5(x)}{H^5(x)}\ , \ \ \  
R(x) = \frac{Q(x)Q(x^5)}{Q^2(x^3)}\ , \\[1mm]
&& {\cal Z}^{1/{\cal N}}= \frac{x^{-1/3}G^3(x)Q^2(x^5)}{H^2(x)}\prod_{n=1}^\infty
\frac{(1-x^{3n-2})(1-x^{3n-1})}
{(1-x^{3n})^2} \ .
\nonumber 
\end{eqnarray}
A nonanalytic behavior of the partition function at 
$z=z_c=(11+5\sqrt{5})/2\approx 11.09017$ as $x\to \pm 1$
indicates presence of a critical point. Such a continuous order-disorder
transition corresponds to a spontaneous occupation of one
of the three triangular sublattices at high densities. Expansion of $\cal Z$ 
in powers of $z-z_c$ yields exact critical exponents 
of the hard-hexagon model, \cite{baxter} 
whereas numerical evaluation of rapidly converging infinite
products provides dependence of various physical
quantities on the chemical potential (temperature).

Low-temperature behavior of  a spin-1/2 kagom\'e  in the vicinity
of the saturation field can be obtained from the above exact solution
by a trivial rescaling from the number of hexagons 
to the number of spins and using the relation $\mu=H_s-H$. 
The entropy as a function of magnetic field at constant $T$ reaches a 
sharp maximum at $H=H_s$. At the saturation field ($z=1$, $x=-0.25496$)
both the entropy and the magnetization have universal 
temperature independent values: 
\begin{equation}
{\cal S}_s/N = 0.11108\approx 0.16026\,\ln 2 \ , \ \ \ \ \ 
M_s/N = 0.44596 \ .
\label{KT0}
\end{equation}
The above value of the magnetization corresponds to 
the density $n_s/N\approx 0.054$ of localized magnons. 
This density well exceeds an estimate $n\sim 0.035$ from the
previous subsection, which signifies suppression of the non-hard-hexagon
states. As an external field is further
decreased there is a continuous crystallization 
transition into a close-packed structure 
(Fig.~\ref{crystal})
with broken translational symmetry.
The transition takes place at a temperature
dependent critical field $H_c(T)$:
\begin{equation}
H_c(T) = H_s - T\ln z_c \approx 3J - 2.40606T \ , 
\label{Hckag}
\end{equation} 
whereas the magnetization has again a universal value:
\begin{equation}
M_c/N = 
1/2 - 1/3\,\rho_c \ , \ \ 
\rho_c = (5-\sqrt{5})/10 \ .
\end{equation} 
The density of magnons at the transition point $n_c/N=\rho_c/3\approx 0.0921$
is about 20\% smaller than the density of the ideal
close-packed structure $n/N=1/9$.
The second order transition at $H_c(T)$ belongs 
to the universality class of a two-dimensional three-state 
Potts model\cite{wu} and has the following
exact critical exponents: $\alpha=1/3$, $\beta=1/9$, $\gamma=13/9$,
and $\nu = 5/6$. \cite{baxter} \ The uniform susceptibility
$\chi=\partial M/\partial H$ also diverges at the transition
point with the same critical exponent 
$\alpha$ as the specific heat. 
Apart from a sharp lambda-peak at $z=z_c$, 
the specific heat has also a rounded Schottky type anomaly
at $z_m=0.03897$ ($x=-0.03296$, $\ln z_m = -3.24502$).

\begin{figure}[t]
\begin{center}
\includegraphics[width=14cm]{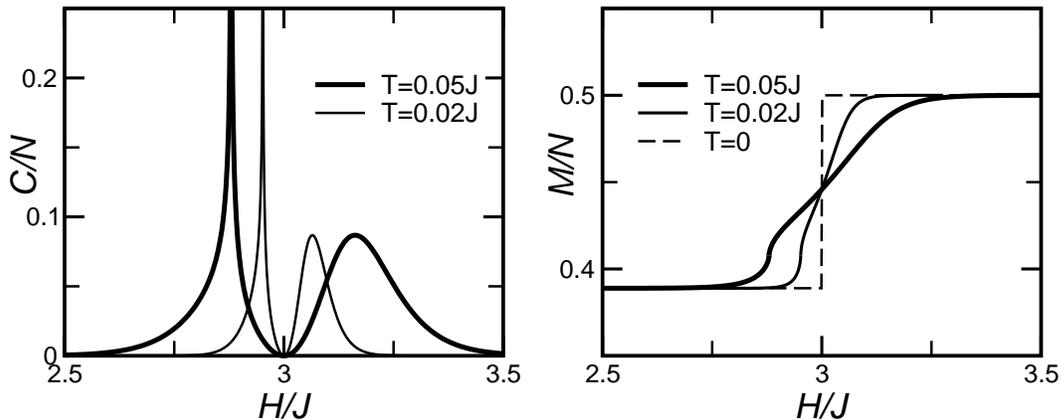}
\end{center}
\caption{\label{Kaghscan} 
Field dependence of the specific heat (left) and the magnetization (right)
of a spin-1/2 kagom\'e antiferromagnet 
obtained in the hard-hexagon model. }
\end{figure}

\begin{figure}[t]
\begin{center}
\includegraphics[width=14cm]{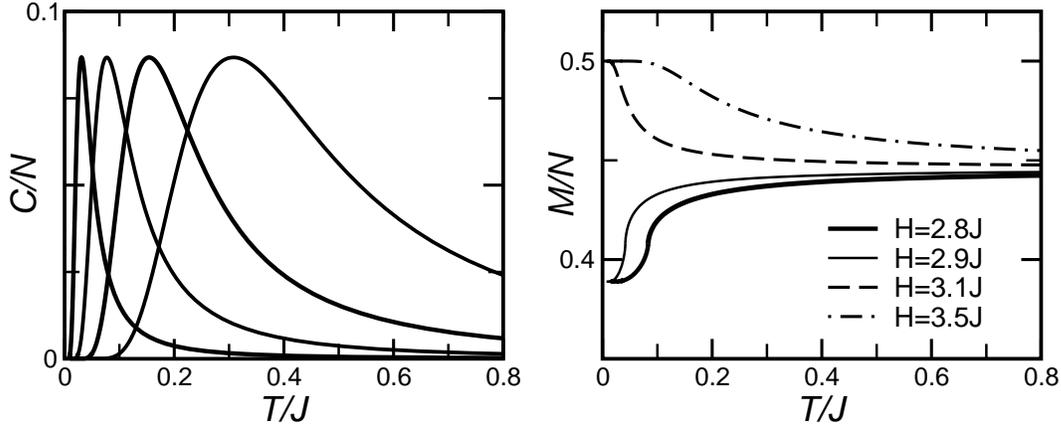}
\end{center}
\caption{\label{Kagtscan} 
Temperature dependence of the specific heat (left) and the magnetization (right)
of a spin-1/2 kagom\'e antiferromagnet 
obtained in the hard-hexagon model. Curves on the left panel correspond
to $H=3.1J$, $3.25J$, $3.5J$, and $4J$ from left to right.}
\end{figure}

Hard-hexagon results for the field dependence of the specific heat 
and the magnetization of a spin-1/2 kagom\'e antiferromagnet 
are shown in Fig.~\ref{Kaghscan}. 
Positions of 
two peaks in the specific heat scale linearly with temperature.
As a function of temperature at a fixed magnetic
field the specific heat  
also exhibits a characteristic low-temperature peak
at $T_m = (H_s-H)/\ln z_m$, see  Fig.~\ref{Kagtscan}.
At higher temperatures $T\sim J$ there should be a second
maximum in $C(T)$, which is not captured by hard-hexagon mapping 
and corresponds 
to development of short-range spin correlations.
Overall, the behavior
of the specific heat and the magnetization of a kagom\'e
antiferromagnet resembles the previous
results for the sawtooth chain. 
The major difference is that
while one-dimensional model has only a finite temperature 
crossover, in two-dimensions the breaking of a discrete
translational symmetry occurs via a sharp second-order transition.
Experimental observations of a double peak structure
in a temperature (field) dependence of
the specific heat and of a characteristic nonmonotonic
temperature dependence of the magnetization 
(Fig.~\ref{Kagtscan}, right panel) could be
a clear signature of degenerate magnons in a geometrically
frustrated magnetic material.

Since the gaps between the noninteracting localized
magnons and propagating states are smaller for a kagom\'e
antiferromagnet, the quantitative validity of the hard-hexagon mapping 
is restricted to lower temperatures $T<0.05$--$0.1J$
and to higher magnetic fields $H>0.9H_s$.
The critical behavior obtained from the exact solution for
hard hexagons should, however, remain valid even beyond
the above range.
At present there are no numerical 
results on the magnetothermodynamics of a spin-1/2 kagom\'e antiferromagnet
in order to compare them with the hard-hexagon mapping.
Note, that exact diagonalization results on
finite periodic clusters should suffer from rather 
strong finite size effects.
They are determined by
a presence of large number of non-hard-hexagon states of localized
magnons, such as the `defect' 
states discussed earlier in this section, whose contribution
vanishes only in the thermodynamic limit $N\to\infty$.

The order parameter of the hard-hexagon model is defined as a difference
of hexagon densities $\rho_i$ on adjacent sites:
$R=\rho_i-\rho_{i+a}$. \cite{baxter} \ Up to a renormalization
prefactor $R$ is equal 
to the Fourier harmonics $\rho_{\bf q}$ 
at ${\bf q}=(4\pi/3a^*,0)$, where $a^*$ is a period
of a triangular lattice. In a kagom\'e antiferromagnet
the magnon crystal has two types of spins, 
on hexagons with localized magnons and between them. 
The two sublattices have different average magnetizations:
$\langle S^z\rangle = 1/3$ and 1/2 at $T=0$.
Hence, the crystalline order of magnons reveals itself
as an extra Fourier harmonics in
the magnetic structure factor $S^{zz}({\bf q})$.
The wave-vector of magnon crystal is the same one as for
a so-called $\sqrt{3}\times\sqrt{3}$ structure, stabilized
in zero field for a kagom\'e antiferromagnet with large
$S\gg 1$. \cite{chubukov} \  Note, that the magnon crystal state
is not reduced to a semiclassical collinear two-sublattice order.
The quantum coherence of spin-flip propagating around one hexagon
plays a crucial role in stabilizing the magnon crystal.

\begin{figure}[t]
\begin{center}
\includegraphics[width=9cm]{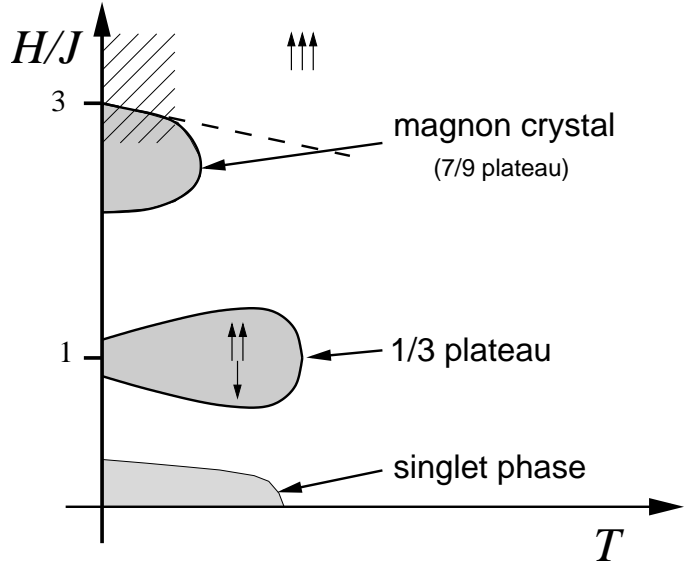}
\end{center}
\caption{
Schematic $H$--$T$ phase diagram of a Heisenberg spin-1/2 kagom\'e
antiferromagnet. Dashed area shows a region of applicability
of the hard-hexagon mapping.}
\label{HTdiag} 
\end{figure}

Finally, the schematic phase diagram of a Heisenberg spin-1/2
kagom\'e antiferromagnet in external
magnetic field is shown in Fig.~\ref{HTdiag}.
In zero magnetic field this quantum frustrated model has a nonmagnetic
ground state with a large number of low-lying
singlet excitations between the ground state and the lowest
excited triplet. \cite{lechem97,waldt98} \ 
Presence or absence of a finite-temperature phase transition
at $H=0$ depends on 
whether such a singlet ground state
is a valence bond crystal, which breaks certain discrete
lattice symmetries, or a true spin-liquid state with
full symmetry of the Heisenberg Hamiltonian.
The answer to this question is not known at present due to
a limited size of clusters accessible for numerical diagonalization.
Applied magnetic field eventually closes the triplet gap 
and the system transforms into a state with finite
magnetization. Numerical investigations of the magnetization curve
of a spin-1/2 Heisenberg model find a clear plateau at
1/3 of the saturation magnetization. \cite{hida,cabra02}\ 
The ground state at the 1/3-plateau has a long-range order of 
valence-bond type with the same $\sqrt{3}\times\sqrt{3}$ periodicity
as the magnon crystal in Fig.~\ref{crystal}. \cite{cabra04}\ 
There is, therefore, a line of phase transitions, which separates
the 1/3 plateau state from a paramagnetic phase and phases
above and below the plateau.
The magnon crystal state
is the third presently known phase of a spin-1/2 kagom\'e antiferromagnet,
which is stabilized in the vicinity of the saturation field.
At zero temperature the magnon crystal
corresponds to the 7/9-magnetization
plateau. The hard-hexagon mapping obtained in Ref.~\citen{mzh04}\ 
and further developed in the present work is valid in the dashed region
of the phase diagram. 
Being restricted to high fields and low temperatures the hard-hexagon
mapping is, nevertheless, quite important, since it provides
the only exact information
about the phase diagram of a spin-1/2 Heisenberg antiferromagnet on a
kagom\'e lattice.

Between the above three phases, magnetization 
of a kagom\'e antiferromagnet grows continuously, \cite{hida,cabra02} which
points at a gapless excitation spectrum.
The number and the nature of such intermediate phases 
remain an open issue.
Gapless states with finite magnetization can also
have nontrivial properties and exotic order
parameters even in the classical limit $S\to\infty$.
\cite{mzh02}\

\section{Discussion}

The above analysis 
of the magnetothermodynamics of the sawtooth chain 
and of the quantum kagom\'e antiferromagnet has been focused
mainly on spin-1/2 models. Apart from a detailed presentation 
of the previous results \cite{schulenburg,schmidt,mzh04}
we have have developed an
effective Ising description 
of localized magnons for the sawtoth chain,
which is valid up to significantly
higher temperatures than the hard-dimer approximation.
Most of the discussed results remain 
valid for an arbitrary quantum spin $S$. These include
the universal values of the entropy at $H=H_s$ and the finite temperature transition
into the magnon crystal state for a kagom\'e lattice model.
The zero temperature magnetization jumps become, however,
increasingly small for large $S$: a relative height of the jump decreases
as $\Delta M = 1/(9S)$ for a kagom\'e antiferromagnet.
The temperature range for stability of the magnon crystal
also becomes smaller with increasing $S$. Indeed, 
the gap between localized magnon states
and higher-energy propagating states is proportional
to the exchange constant $J$.
(Variational calculations for the $S=1$ sawtooth chain
yield only a slightly different gap $\Delta=0.41J$ compared to 
the $S=1/2$ result of Sec.~3.)
The crystallization temperature scales, therefore, as $T_c=O(J)$
and becomes a small
fraction of the Curie-Weiss constant 
$T_c/\theta_{CW}\sim 1/S^2$.
The domain of quantum effects is pushed
to extremely low temperatures for semiclassical spins $S\gg 1$ 

A strong magnetocaloric effect discussed in Sec.~3 remains,
however, present even in the classical limit $S\to\infty$.
\cite{mzh03}\  Recently, a large temperature decrease during
an adiabatic demagnetization process has been measured in
pyrochlore compound Gd$_2$Ti$_2$O$_7$ with $S=7/2$.
\cite{sosin}\ Note, 
that antiferromagnets on a pyrochlore and a checkerboard lattices 
also have localized magnon excitations. 
Their low-temperature behavior
is, therefore, mapped to lattice gas models
of hard-core classical particles. 
For the checkerboard antiferromagnet an appropriate
low-temperature model is a square lattice gas of hard-core particles with
nearest- and next-nearest neighbor exclusions.
There is no exact solution for such statistical model.
Still, effective representations may be quite useful,
because lattice gas models can be efficiently studied with a powerful
classical Monte Carlo method.
The thermodynamics of the corresponding lattice models generally depends on a 
dimensionless fugacity $z=e^{(H_s-H)/T}$. This immediately implies
a strong magnetocaloric effect $T\to 0$ as $H\to H_s$ for all
frustrated magnets with localized spin-flips.
The residual entropy at $H=H_s$, which determines a cooling
power, does depend on a lattice
geometry. An interesting open problem is calculation of
the residual entropies for pyrochlore and checkerboard
antiferromagnets in order to determine the  favorable 
geometrically frustrated spin system for applications
in adiabatic refrigerators. \cite{lounasmaa,refrig}\ There 
is also a certain difference in the magnetocaloric effect
for quantum and (semi)classical frustrated magnets.
The quantum spin systems exhibit a strong cooling on the two sides
of the saturation field when $H\to H_s\pm 0$: both 
the saturated phase at $H>H_s$ and the magnon crystal
at $H<H_s$ have no entropy at $T=0$.
Frustrated classical spin system has in contrast infinite degeneracy
of the ground state for $H<H_s$. Therefore, a much stronger cooling
effect is found when applied field is decreased towards $H_s$
from the high-field side. \cite{mzh03}\ For
intermediate values of spin $S>1$ the quantum order by disorder effect
\cite{shender} lifts
the classical degeneracy at $H<H_s$ and restores zero entropy.
This leads to an asymmetric
field dependence for $T_{\cal S}(H)$ with the steepest slope
at $H>H_s$ and the lowest temperature
achievable in the vicinity of the saturation field. 

Real magnetic materials usually have additional weak interactions
apart from a nearest-neighbor Heisenberg exchange.
These include, for example, various an\-isot\-ropies, next-nearest-neighbor exchanges,
and the dipolar interactions. Such extra interactions lift
the classical degeneracy of frustrated magnets and in the
quantum case induce a weak dispersion of the lowest magnon branches
(\ref{wkag}) and (\ref{wsaw}). 
Phase transition at $H=H_s$ corresponds, then, to a usual condensation
of a single magnon mode at a certain wave-vector $\bf Q$.
The finite zero-temperature entropy at the saturation field is completely destroyed in
such a case.
The question, whether some of the effects described in the present
paper are still observable or not, can be solved by a simple comparison
of the energy scales. If the gap in a magnon crystal state exceeds
the bandwidth $W$ of the lowest branch $\omega_1({\bf k})$, then
there has to be a subsequent phase transition 
between a state with transverse
magnetic order in the immediate vicinity of $H_s$ and  
the crystal state without such ordering. (There will be, of course,
zero-point fluctuations in the magnon crystal ground state 
on top of a simple product of localized magnon wave-functions,
Fig.~\ref{crystal}.)
As long as the energy scale for weak residual interactions is
significantly smaller than the nearest-neighbor exchange $W\ll J$,
a frustrated magnet continues to exhibit a large magnetocaloric
effect. The lowest reachable temperature is, however, limited to
$T_{\rm min}\sim W$.
Finally, one of the suitable ways to detect experimentally 
nearly dispersionless magnon bands in frustrated
magnets is to measure a low-temperature peak
in the specific heat for 
$T_m=(H-H_s)/\ln z_m >W$.

\section*{Acknowledgments}

We are grateful to A. Honecker for providing us the numerical
data for the sawtooth chain. H.\,T.\ is supported by 
a Grant-in-Aid from the Ministry of
Education, Science, Sports, and Culture of Japan.


\begin{thebibliography}{99}


\bibitem{ramirez02}
A. P. Ramirez, B. S. Shastry, A. Hayashi,  J. J. Krajewski, D. A.
Huse, and R. J. Cava, Phys.\ Rev.\ Lett.\ \textbf{89} (2002), 067202. 

\bibitem{hov}
S. Hov, H. Bratsberg, and A. T. Skjeltorp,
J. Magn. Magn. Mater. {\bf 15--18} (1980), 455.

\bibitem{takagi}
H. Ueda, H. A. Katori, H. Mitamura, T. Goto, and H. Takagi,
Phys. Rev. Lett. {\bf 94} (2005), 047202. 

\bibitem{schulenburg}
J. Schulenburg, A. Honecker, J. Schnack, J. Richter, and H.-J. Schmidt,
Phys. Rev. Lett. {\bf 88} (2002), 167207.

\bibitem{schnack}
J. Schnack, H.-J. Schmidt, J. Richter, and J. Schulenburg,
Eur. Phys. J. B {\bf 24} (2001), 475.

\bibitem{schmidt}
H.-J. Schmidt, J. Phys. A {\bf 35} (2002), 6545.

\bibitem{mzh04}
M. E. Zhitomirsky and H. Tsunetsugu, Phys. Rev. B {\bf 70} (2004), 100403.

\bibitem{richter}
J. Richter, J. Schulenburg, A. Honecker, J. Schnack, and H.-J. Schmidt,
J. Phys.:\ Condens.\ Matter {\bf 16} (2004), S779.

\bibitem{1d}
M. E. Zhitomirsky and A. Honecker,
J. Stat. Mech.: Theor. Exp. (2004), P07012. 

\bibitem{derzhko}
O. Derzhko and J. Richter, Phys. Rev. B {\bf 70} (2004), 104415.

\bibitem{mattis}
D. C. Mattis, \textit{The Theory of Magnetism I} (Springer, Berlin, 1981).

\bibitem{batyev} E. G. Batyev and L. S. Braginskii,
Zh. \'Eksp. Teor. Fiz. {\bf 87} (1984), 1361 
[Sov. Phys. JETP {\bf 60} (1984), 781].

\bibitem{jackeli}
G. Jackeli and M. E. Zhitomirsky, Phys. Rev. Lett. {\bf 93} 
(2004), 017201.

\bibitem{scgo}
A. P. Ramirez, G. P. Espinosa and A. S. Cooper, 
Phys. Rev. B {\bf 45} (1992), 2505. 

\bibitem{delafossite}
K. Isawa, Y. Yaegashi, M. Komatsu, M. Nagano, S. Sudo, M. Karppinen, 
and H. Yamauchi,
Phys. Rev. B {\bf 56} (1997), 3457.

\bibitem{mzh03}
M. E. Zhitomirsky, Phys. Rev. B {\bf 67} (2003), 104421.

\bibitem{metcalf}
B. D. Metcalf and C. P. Yang, Phys. Rev. B {\bf 18} (1978), 2304.

\bibitem{baxter}
R. J. Baxter, \textit{Exactly Solved Models in Statistical Mechanics}
(Academic Press, London, 1982);
R. J. Baxter, J. Phys. A {\bf 13} (1980), L61;
R. J. Baxter and S. K. Tsang, J. Phys. A {\bf 13} (1980), 1023;
R. J. Baxter and P. A. Pearce, J. Phys. A {\bf 15} (1982), 897.

\bibitem{wu}
F. Y. Wu, Rev. Mod. Phys. {\bf 54}, (1982), 235.

\bibitem{chubukov}
A. Chubukov, Phys. Rev. Lett. {\bf 69} (1992), 832.

\bibitem{lechem97}
P. Lecheminant, B. Bernu, C. Lhuillier, L. Pierre, and P. Sindzingre,
Phys. Rev. B {\bf 56} (1997), 2521.

\bibitem{waldt98}
Ch. Waldtmann, H.-U. Everts, B. Bernu, C. Lhuillier, P. Sindzingre,
P. Lecheminant, and L. Pierre, 
Eur. Phys. J. B {\bf 2} (1998), 501.

\bibitem{hida}
K. Hida, J. Phys. Soc. Jpn. {\bf 70} (2001), 3673.

\bibitem{cabra02}
D. C. Cabra, M. D. Grynberg, P. C. W. Holdsworth, and P. Pujol,
Phys. Rev. B {\bf 65} (2002), 094418.

\bibitem{cabra04}
D. C. Cabra, M. D. Grynberg, P. C. W. Holdsworth, A. Honecker,
P. Pujol, J. Richter, D. Schmalfu\ss, and J. Schulenburg,
Phys. Rev. B {\bf 71} (2005), 144420.

\bibitem{mzh02}
M. E. Zhitomirsky, Phys. Rev. Lett. {\bf 88} (2002), 057204.

\bibitem{sosin}
S. S. Sosin, L. A. Prozorova, A. I. Smirnov, A. I. Golov, I. B. Berkutov, 
O. A. Petrenko, G. Balakrishnan, M. E. Zhitomirsky, 
Phys. Rev. B {\bf 71} (2005), 094413.

\bibitem{lounasmaa}
O. V. Lounasmaa, \textit{Experimental Principles and Methods Below 1K,
chapter 5} (Academic Press, London, 1974).

\bibitem{refrig}
J. A. Barclay and W. A. Steyert, Cryogenics {\bf 22} (1982), 73.

\bibitem{shender}
E. F. Shender, Zh. \'Eksp. Teor. Fiz. {\bf 83} (1982), 326
[Sov. Phys. JETP {\bf 56} (1982), 178].


\end{thebibliography}
\end{document}